\documentclass[mathpazo]{aamm}

\renewcommand\thisnumber{x}
\renewcommand\thisyear {200x}
\renewcommand\thismonth{xxx}
\renewcommand\thisvolume{xx}

\usepackage{float}
\usepackage{dcolumn}
\usepackage{hyperref}\hypersetup{colorlinks=true, allcolors=blue}

\newcommand\Go{G\"{o}rtler }
\newcommand\ie{i.e.,\ }
\newcommand\etal{et al.,\ }
\newcommand{\Rey}   {\mbox{\textit{Re}}}
\newcommand{\Ma}    {\mbox{\textit{Ma}}}
\newcommand{\Prl}   {\mbox{\textit{Pr}}}
\newcommand{\dd}    {\mathrm{d}}

\graphicspath{{Figures/}}
\begin{document}


\markboth{J. Ren, Y. Xi and S. Fu}{The new mode of instability in viscous high-speed boundary layer flows}
\title{The new mode of instability in viscous high-speed boundary layer flows}

\author{Jie Ren\affil{1}\comma\affil{2}, Youcheng Xi\affil{1} and Song Fu\affil{1}\comma\corrauth}
\address{\affilnum{1}\ School of Aerospace Engineering, Tsinghua University, Beijing, 100084, China\\
\affilnum{2}\ Process and Energy Laboratory, Delft University of Technology, Leeghwaterstraat 39, 2628 CB Delft, The Netherlands}
 \email{\tt fs-dem@tsinghua.edu.cn}
%
%
%

\begin{abstract}
The new mode of instability found by Tunney \etal \cite{Tunney2015} is studied with viscous stability theory in this article. When the high-speed boundary layer is subject to certain values of favorable pressure gradient and wall heating, a new mode becomes unstable due to the appearance of the streamwise velocity overshoot ($U(y)>U_\infty$) in the base flow. The present study shows that under practical Reynolds numbers, the new mode can hardly co-exist with conventional first mode and Mack's second mode. Due to the requirement for additional wall heating, the new mode may only lead to laminar-turbulent transition under experimental (artificial) conditions.
\end{abstract}

\keywords{boundary layer stability, compressible boundary layers, high-speed flow.}

\ams{65M10, 78A48}

\maketitle

\section{Introduction}\label{S1}
The mechanism of high-speed laminar-turbulent flow transition is far from fully understood \cite{Fedorov2011}. One important reason is the multitudinous routes of the transition process that is in turn influenced by various environmental conditions. Among them, modal stability is generally considered the fundamental mechanism and relatively well-studied. The representative examples are Tollmien-Schlichting waves in (quasi-) parallel flows \cite{BL}, Mack's second modes in hypersonic flows \cite{Mack1984}, cross-flow modes in three-dimensional boundary layers \cite{Saric2003} and \Go modes over concave surfaces (when Reynolds number is large) \cite{Saric1994}. Under certain conditions (particularly with low external turbulence and smooth geometry), perturbations (generated through receptivity mechanism) get amplified with modal instabilities causing the flow close to transition when their amplitude becomes large. However, even after amounts of studies, the knowledge on this fundamental modal stability is still insufficient.

Compared with zero pressure gradient, favorable pressure gradient (hereafter referred to as FPG) significantly stabilizes the boundary layer in both incompressible and compressible flows (the first mode as well as Mack's second mode). This is supported by a number of studies, e.g., with direct numerical simulation \cite{Kloker1990,Bech1998,Franko2014}, linear stability theory \cite{Malik1989,Zurigat1992,Masad1994} and very recent experiments \cite{Tokugawa2015,Costantini2016}. Hence, in the review by Reed et al.\cite{Reed1996}, the instability of boundary layer with FPG is described as ``\emph{very weak, if it exists at all}''. In fact, with FPG, the profile of the base flow $U(y)$ becomes fuller, and the thickness of the boundary layer is decreased, which is mainly responsible for the stabilization of the boundary layer.

On the other hand, wall-heating/cooling is one of the common passive flow control methods used on various occasions. Its influence on boundary layer stability has been well documented (see reviews in \cite{Mack1987,Reed1996}). In contrast to the adiabatic condition, wall heating can destabilize the first mode while stabilizing Mack's second mode. Wall cooling, instead, has opposite effects. One shall distinguish between wall-heating and localized wall-heating. The latter gives rise to wall temperature jump effect and can destabilize Mack's second mode (see recent analysis in \cite{Fedorov2015}).

When the flow is subject to the dual effects of FPG and wall-heating, a new mode comes to light. A first analytical study was performed by Tunney \etal \cite{Tunney2015} under the inviscid assumption. The direct cause of the instability is the appearance of streamwise velocity \textit{overshoot} ($U(y)>U_\infty$ near the upper edge of the boundary layer). Discussion on the \textit{overshoot} can be found in Tunney \etal \cite{Tunney2015} and the references therein. Under inviscid assumption, the new mode was shown to have comparable growth rate as the conventional first mode and Mack's higher mode. However, the possible importance of the new mode is not evaluated. In this paper, we report a viscous stability analysis (with spatial mode) on this new mode which is more relevant for developing boundary layers. The impact and limitations of the new mode will be discussed. In Section~\ref{S2} the methodology and the base flow are introduced. Modal stability is discussed in Section~\ref{S3}, and the paper is concluded in Section~\ref{S4}.

\section{Methodology and base flow}\label{S2}
The stability equations are derived from the Navier-Stokes equations provided the base flow is obtained in advance. A frequently-adopted form is written as
    \begin{multline}\label{ES0}
    {\boldsymbol{\Gamma }}\frac{{\partial {{\tilde q}}}}{{\partial t}}+ {\boldsymbol{A}}\frac{{\partial {{\tilde q}}}}{{\partial x}} +
    {\boldsymbol{B}}\frac{{\partial {{\tilde q}}}}{{\partial y}} +
    {\boldsymbol{C}}\frac{{\partial {{\tilde q}}}}{{\partial z}} +
    {\boldsymbol{D}{\tilde q}} \\=
    {{\boldsymbol{V}}_{xx}}\frac{{{\partial ^2}{{\tilde q}}}}{{\partial {x^2}}} + {{\boldsymbol{V}}_{xy}}\frac{{{\partial ^2}{{\tilde q}}}}{{\partial x\partial y}} + {{\boldsymbol{V}}_{xz}}\frac{{{\partial ^2}{{\tilde q}}}}{{\partial x\partial z}} + {{\boldsymbol{V}}_{yy}}\frac{{{\partial ^2}{{\tilde q}}}}{{\partial {y^2}}} + {{\boldsymbol{V}}_{yz}}\frac{{{\partial ^2}{{\tilde q}}}}{{\partial y\partial z}} + {{\boldsymbol{V}}_{zz}}\frac{{{\partial ^2}{{\tilde q}}}}{{\partial {z^2}}}.
    \end{multline}
Here $\tilde q=(\tilde \rho, \tilde u, \tilde v, \tilde w, \tilde T)^T$ is the perturbation vector of flow density, velocity and temperature. The $5\times5$ matrices ${\boldsymbol{\Gamma }}$, ${\boldsymbol{A}}$, ${\boldsymbol{B}}$... are functions of the base flow and dimensionless parameters \Rey, \Ma, \Prl. Detailed expressions for these matrices can be found in the authors' previous articles \cite{RJ2014b,RJ2015b}. The physical quantities are nondimensionalized with their corresponding free-stream values except pressure $p^*$ by $\rho^*_\infty U^{*2}_\infty$. Asterisk denotes dimensional quantities. The orthogonal coordinates $x^*$, $y^*$, $z^*$ describing the distance in streamwise, normalwise and spanwise directions are normalised with the local boundary layer thickness length scale $\delta^*=\sqrt{\nu^*_\infty x^*/U^*_\infty}$. As a result, the dimensionless parameters \Rey, \Ma, \Prl~are
    \begin{equation}
    \Rey=\dfrac{\rho^*_\infty U^*_\infty \delta^*}{\mu^*_\infty},~~~
    \Ma=\dfrac{ U^*_\infty}{\sqrt{\gamma R^*_{air}T^*_\infty}},~~~
    \Prl=\dfrac{\mu^*_\infty C^*_p}{\kappa^*_\infty}.
    \end{equation}
One is able to identify, \Rey~is a measure of streamwise coordinate when the freestream parameters are fixed. On the other hand, when $\Rey\rightarrow\infty$, the equations reduce to inviscid O-S and Squire equations in compressible form. In the framework of modal stability, equation \eqref{ES0} is solved as an eigenvalue problem through
    \begin{equation}
    {\tilde q}(x,y,z,t)=\hat{q}(y)\exp(i\alpha x+i\beta z-i\omega t)+c.c.
    \end{equation}
We focus on the spatial problem which is more relevant to practical boundary layer flows. Therefore, $\alpha$ is the eigenvalue to be numerically solved. In the above formulation, we have assumed the fluid to be calorically-perfect-gas and \Prl~is constant. Therefore,
    \begin{equation}
    p^*=\rho^*R^*_{air}T^*,~~
    \gamma=1.4,~~
    C^*_p=\text{const},~~
    R^*_{air}=\text{const},~~
    \Prl=0.72=\text{const}.
    \end{equation}
The first coefficient of viscosity $\mu$ is given by Sutherland's law and the second coefficient follows Stokes's hypothesis, \ie $\lambda=-2/3\mu$. The code is carefully validated with published results \cite{RJ2014b,RJ2016}, one example is also provided in Fig.~\ref{F4}(\textit{a}).

The self-similar solution of the boundary layer equations offers a concise thus normalized base flow. For a better understanding of the new mode and generation of the full stability diagram, it is employed in this study. Introducing the Mangler-Levy-Lees transformation (see detailed introduction in \cite{BL,Cebeci1974,Cebeci2002})
    \begin{equation}\label{ET}
    \left.
    \begin{aligned}
    {\rm d}\xi = \rho_e\mu_eu_e~\dd x\\
    {\rm d}\eta=\frac{\rho u_e}{\sqrt{2\xi}}~\dd y
    \end{aligned}
    \right\}
    \end{equation}
into the boundary layer equations, yields the transformed equations:
    \begin{equation}\label{EB1}
    \left.
    \begin{aligned}
    (cf'')'+ff''+\beta_p(1+k)(g-f'^2)&=0\\
    (a_1g'+a_2f'f'')'+fg'&=0
    \end{aligned}
    \right\}
    \end{equation}
where the prime denotes the derivative with respect to $\eta$. The coefficients are defined as
    \refstepcounter{equation}
    $$
      c=\frac{\mu}{T}, \quad
      a_1=\frac{c}{\Prl}, \quad
      a_2=\frac{2k}{k+1}\left(1-\frac{1}{\Prl}\right)c, \quad
      k=\frac{(\gamma-1)}{2}\Ma^2, \quad
      \beta_p=\frac{2\xi}{u_e}\frac{\dd u_e}{\dd\xi}.
      \eqno{(\theequation{\mathit{a}-\mathit{e}})}
    $$
The physical quantities are recovered through
    \refstepcounter{equation}
    $$
      \frac{u^*}{u^*_\infty}=f', \quad
      \frac{H^*}{H^*_\infty}=g, \quad
      \frac{T^*}{T^*_\infty}=(1+k)g-kf'^2.
      \eqno{(\theequation{\mathit{a}-\mathit{c}})}.
    $$
Here $H$ denotes the total enthalpy. It should be noted that the temperature-based energy equation is also frequently used. With the same transformation \eqref{ET}, the temperature-based energy equation becomes
    \begin{equation}\label{EB2}
    \frac{1}{\Prl}(c\theta')'+f\theta'+(\gamma-1)\Ma^2cf''^2=0,
    \end{equation}
where
    \begin{equation}
    \frac{T^*}{T^*_\infty}=\theta.
    \end{equation}
\eqref{EB1} or \eqref{EB2} can be solved with standard boundary value problem (BVP) solvers. The boundary conditions (isothermal) are
    \refstepcounter{equation}
    $$
      f(0)=f'(0)=0, \quad
      g(0)=H_w, \quad
      f'(\infty)=g(\infty)=1
      \eqno{(\theequation{\mathit{a}-\mathit{c}})}
    $$
and
    \refstepcounter{equation}
    $$
      f(0)=f'(0)=0, \quad
      \theta(0)=T_w, \quad
      f'(\infty)=\theta(\infty)=1.
      \eqno{(\theequation{\mathit{a}-\mathit{c}})}
    $$
When $g'(0)=0$ or $\theta'(0)=0$ is applied instead of the Dirichlet condition, the flow is adiabatic.

\begin{figure}
    \centering
    \includegraphics[width=0.45\linewidth]{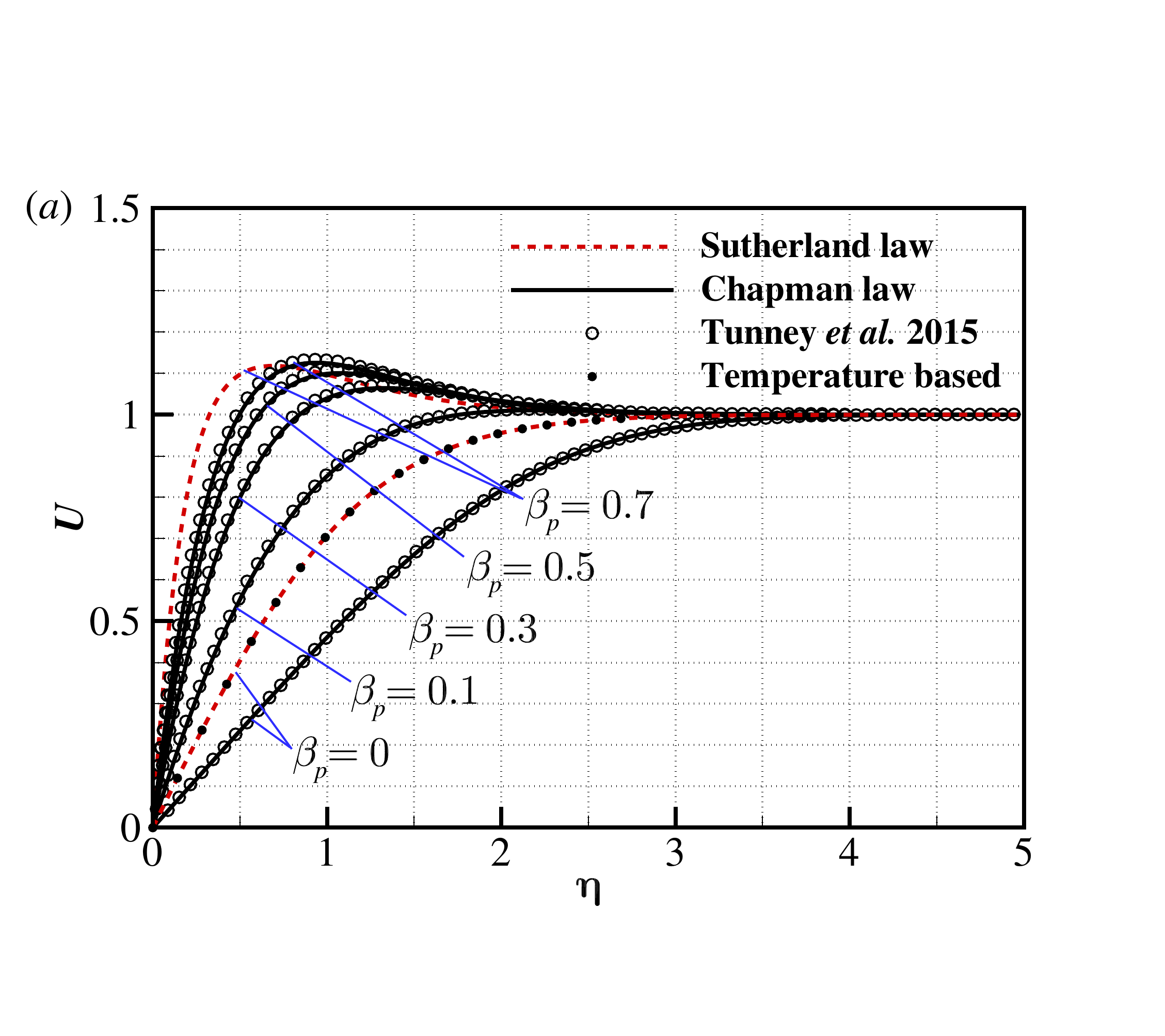}
    \includegraphics[width=0.45\linewidth]{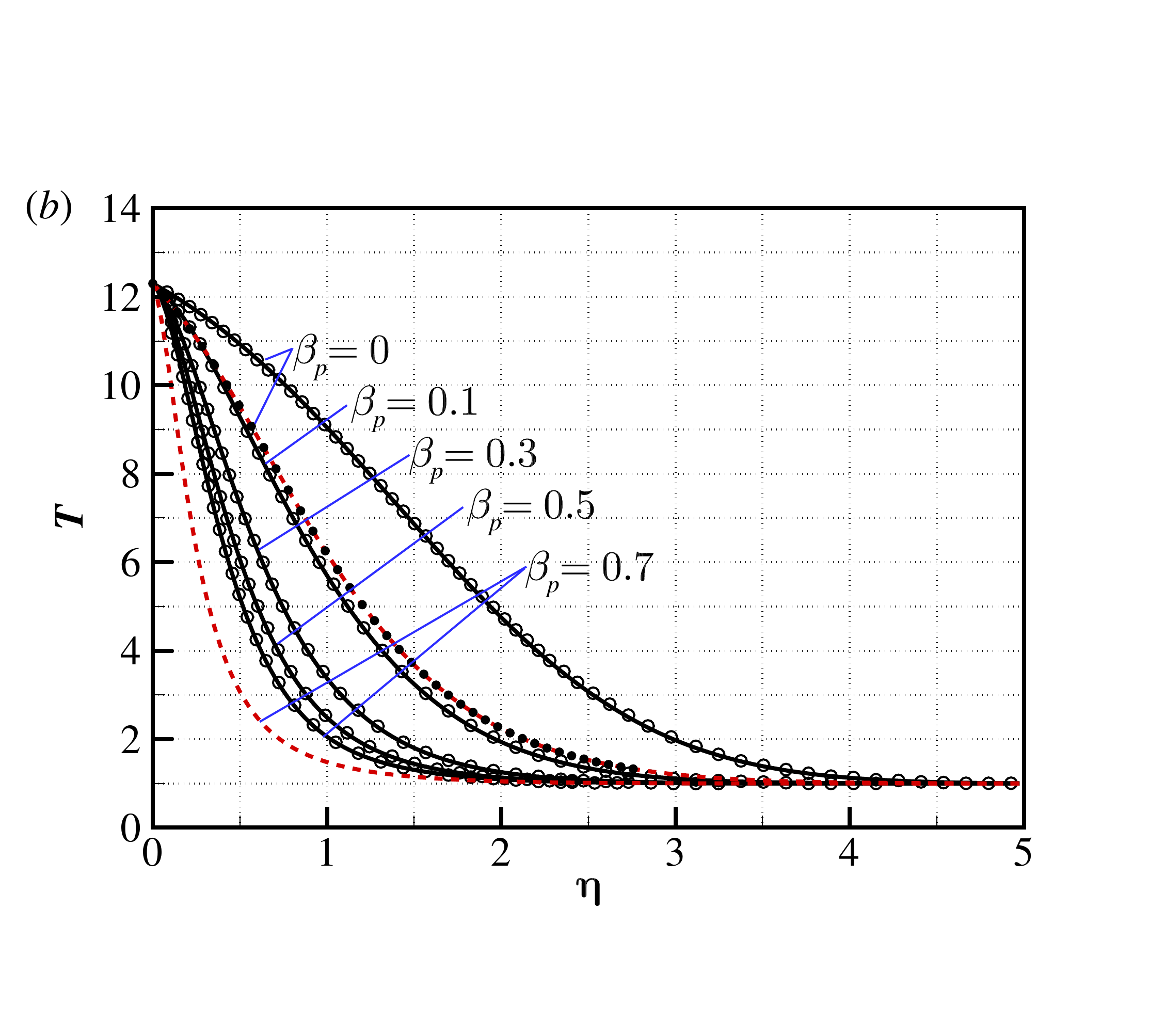}
    \caption{Profiles of (\textit{a}) the streamwise velocity $U$ and (\textit{b}) temperature $T$ as functions of the similarity variable $\eta$. The Falkner-Skan pressure gradient parameter $\beta_p=0,~0.1,~0.3,~0.5$ and $0.7$ respectively. $\Ma=6$, $H_w=1.5$ and $\Prl=0.72$.}
    \label{F1}
    \end{figure}

Figure~\ref{F1} shows the profiles of streamwise velocity and temperature. As can be observed, perfect matches with \cite{Tunney2015}(with Chapman's law) and temperature based energy equation (see also \cite{Ricco2009} with Sutherland's law) have been achieved. Along with the increase of the pressure gradient $\beta_p$, the laminar boundary layer profile $U(y)$ is essentially modified. The boundary layer thickness decreases. An inflection point appears along with the presence of the streamwise velocity overshoot (larger than the free-stream value). And two generalized inflection points are found due to the appearance of velocity overshoot see figure 2 in Tunney \etal \cite{Tunney2015}.

\section{Stability analysis}\label{S3}
Three groups of cases have been studied to reveal the stability diagram of the new mode and its relationship to conventional first and Mack's second mode. See Table \ref{T1} for the prescribed parameters. Case 1 serves as a basic case to recover the typical zero-pressure gradient boundary layer with adiabatic boundary condition. Wall heating is included in Case 2, and the dual effects of wall heating and FPG are considered in Case 3. A broad range of parameters is specified to show all the possible modal instabilities thus allowing a complete stability diagram.

 \begin{table}
    \begin{tabular}{cccc}
    \hline
      Case  & Wall (total) enthalpy &  Wall temperature & Pressure gradient \\
       1    & $H'_w=0$ ($H_w=0.88$)      & $T_w=4.44$ & $\beta_p=0$   \\
       2    & $H_w=1.5$     & $T_w=7.58$ &  $\beta_p=0$   \\
       3    & $H_w=1.5$     & $T_w=7.58$ &  $\beta_p=0.4$ \\
       \hline
    \end{tabular}
    \caption{Parameters of the three cases studied. Mach number $\Ma=4.5$, Stagnation temperature $T_0^*=329K$, Spanwise wavenumber $0\leqslant\beta\leqslant1$, angular frequency $0\leqslant\omega\leqslant1.2$ and Reynolds number $100\leqslant\Rey\leqslant2000$.}
    \label{T1}
    \end{table}

    \begin{figure}
    \centering
    \includegraphics[width=0.48\linewidth]{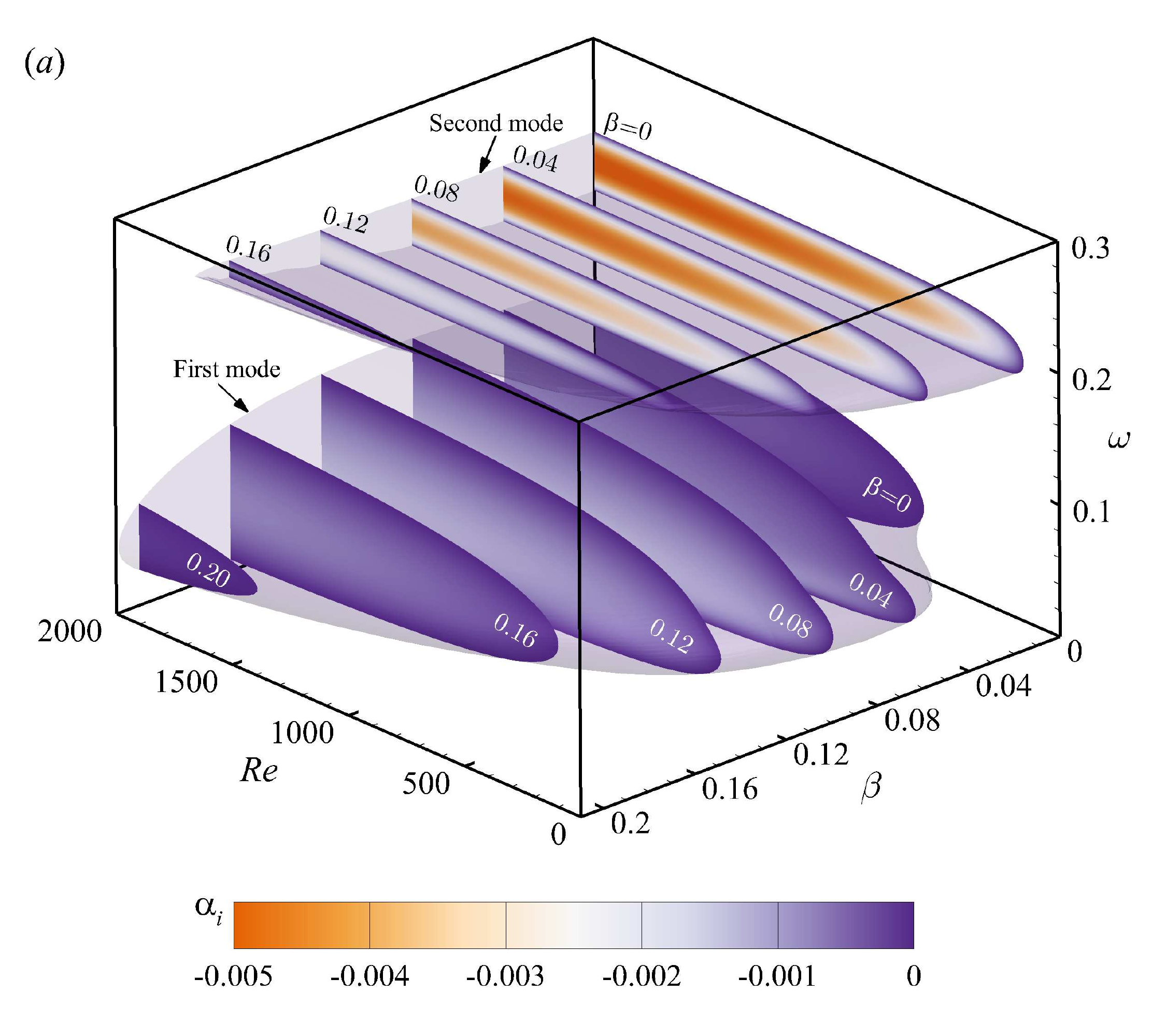}
    \includegraphics[width=0.48\linewidth]{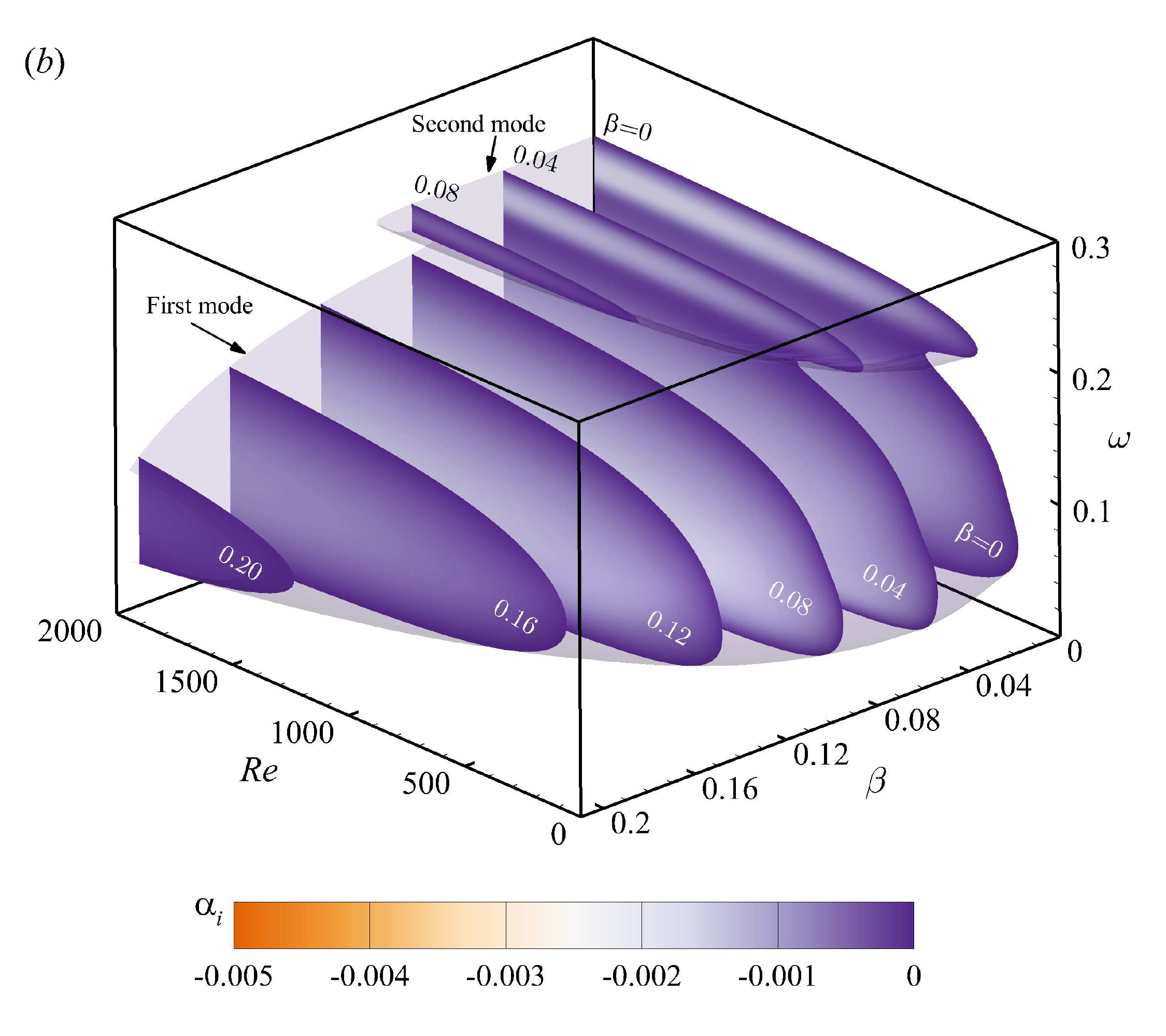}\\
    \includegraphics[width=0.48\linewidth]{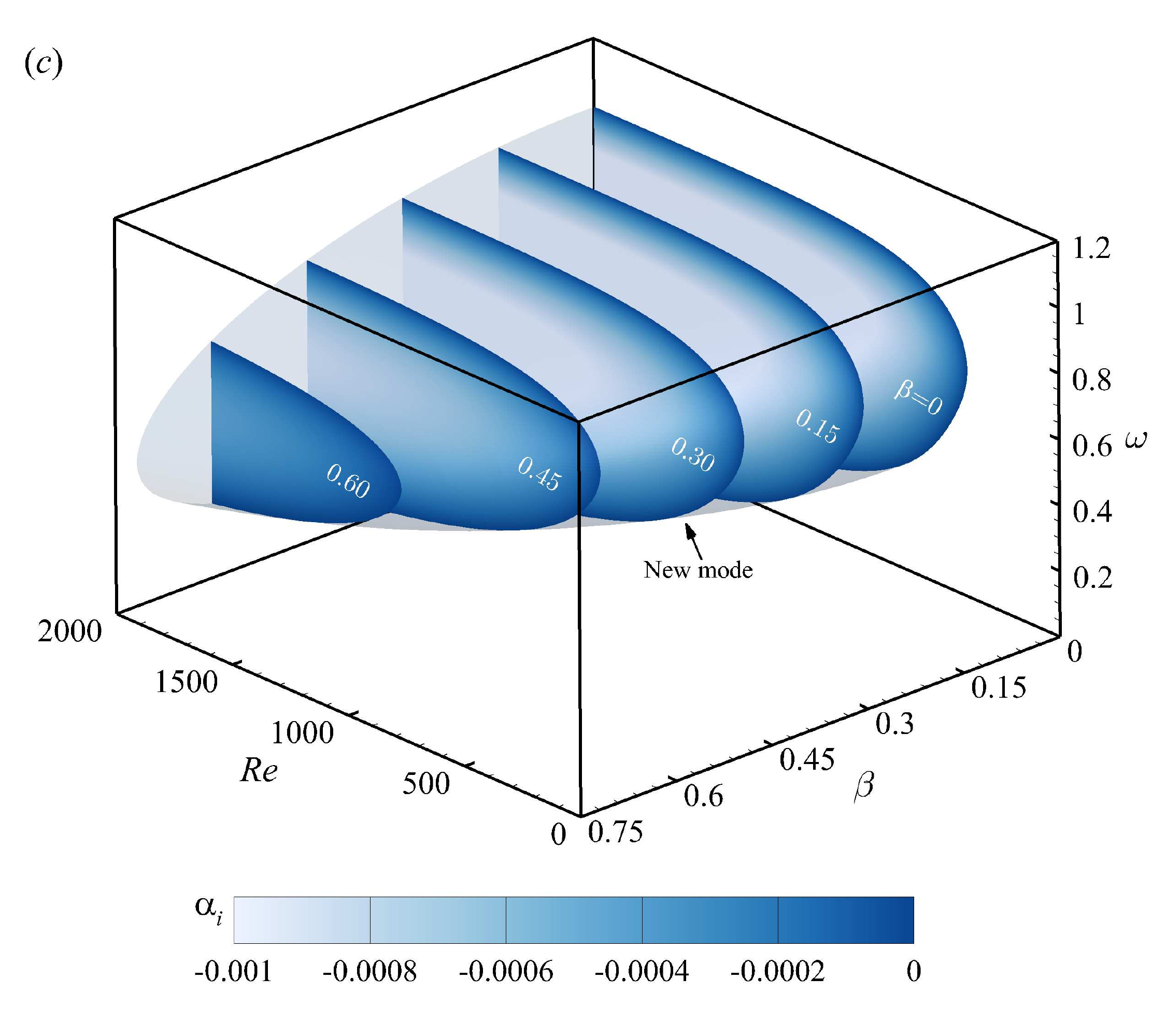}
    \includegraphics[width=0.48\linewidth]{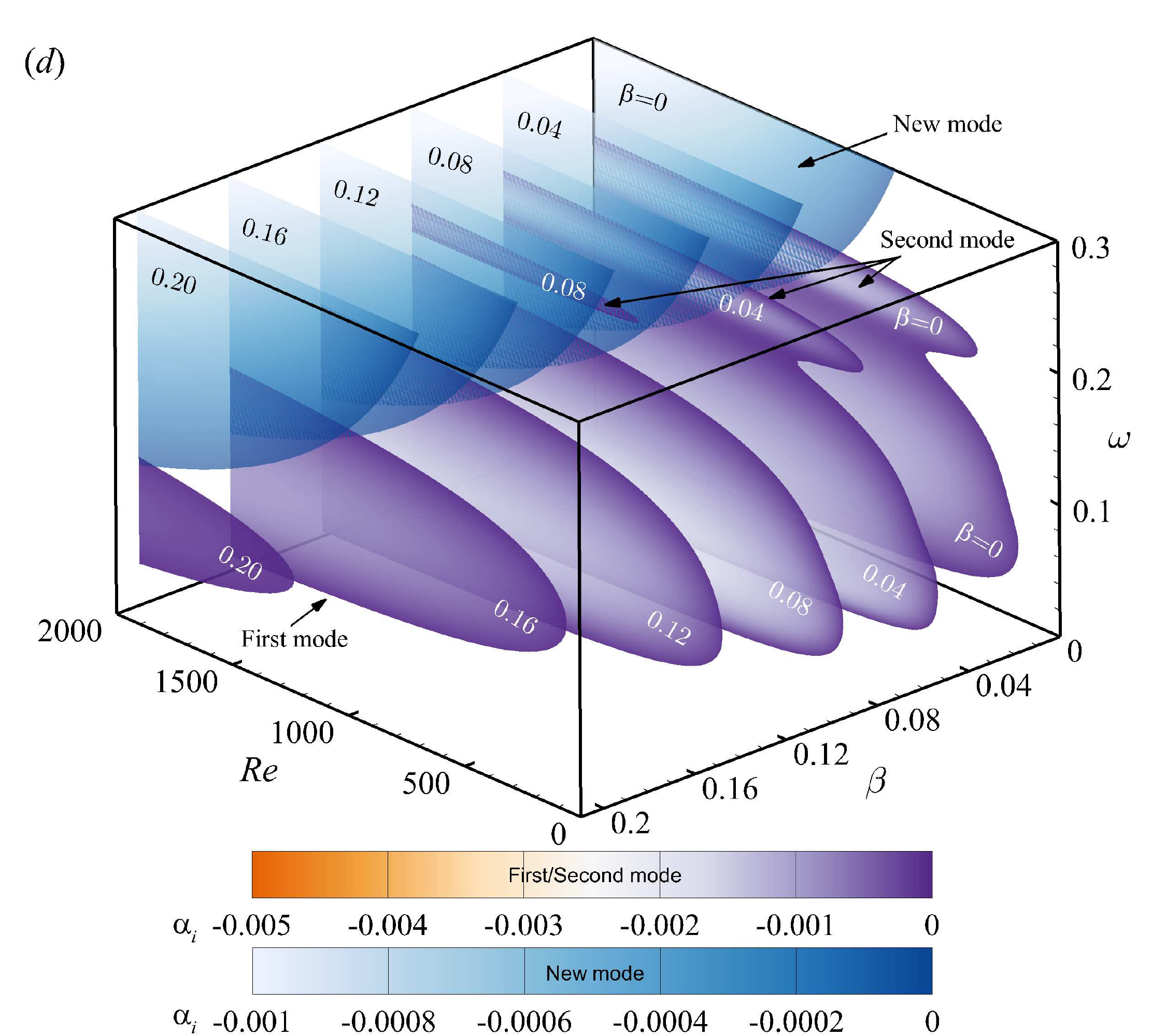}
    \caption{Stability diagram of the boundary layer for Case 1 (\textit{a}); Case 2 (\textit{b}); Case 3 (\textit{c}); Case 2 \& 3 (\textit{d}).}
    \label{F2}
    \end{figure}

Figure~\ref{F2}(\textit{a}) shows the stability diagram of zero-pressure gradient boundary layer subjected to adiabatic boundary conditions (Case 1). The unstable block (in the $\Rey-\beta-\omega$ space) of the first and second modes are enclosed by the corresponding enveloping surfaces. Apparently, both modes become unstable starting from certain $\Rey$ numbers. These numbers, termed critical Reynolds numbers, indicate that the perturbations gain exponential eigen-growth downstream of the leading edge. As can be seen from Figure~\ref{F2}(\textit{a}), the unstable regions of the two modes do not intersect with each other at $\Ma=4.5$. The angular frequency of the second mode is above the first mode, therefore possessing a higher frequency.

Several surface cuts are shown with $\beta=0,~0.04,~0.08,~0.12,~0.16$ and 0.20 respectively. These iso-surfaces show contours of the eigenvalue $\alpha_i$ ($-\alpha_i$ is the local growth rate). One is able to see that the second mode have an obviously larger growth rate and it reaches maximum growth rate at $\beta=0$. When $\beta$ is increased, both the maximum growth rate and the unstable area get reduced. As a result, it is generally accepted that the 2-D perturbation ($\beta=0$) is the most dangerous for the second mode. On the other hand, the optimal spanwise wavenumber for the first mode is not zero. 

When wall-heating is imposed, as shown in Figure~\ref{F2}(\textit{b}), Mack's second mode is significantly stabilized. Both the maximum growth rate and the unstable area become reduced. On the other hand, the first mode is enhanced. The maximum growth rate is not much increased, but the unstable region is expanded to a major degree, intruding into Mack's second mode.

As can be seen in Figure~\ref{F2}(\textit{c}), with the dual effects of wall-heating and FPG, the new mode becomes the only unstable mode in the boundary layer. By comparing with Case 1 and Case 2, the new mode has a much larger unstable region in terms of $\beta$ and $\omega$. Interestingly, it reaches maximum growth rate at $\beta=0$ but has smaller growth rate compared with the conventional modes. Case 2 and 3 are plotted together in Figure~\ref{F2}(\textit{d}). It is apparent that the new mode covers the frequency band of Mack's second mode and extends to much higher values.

    \begin{figure}
    \centering
    \includegraphics[width=0.48\linewidth]{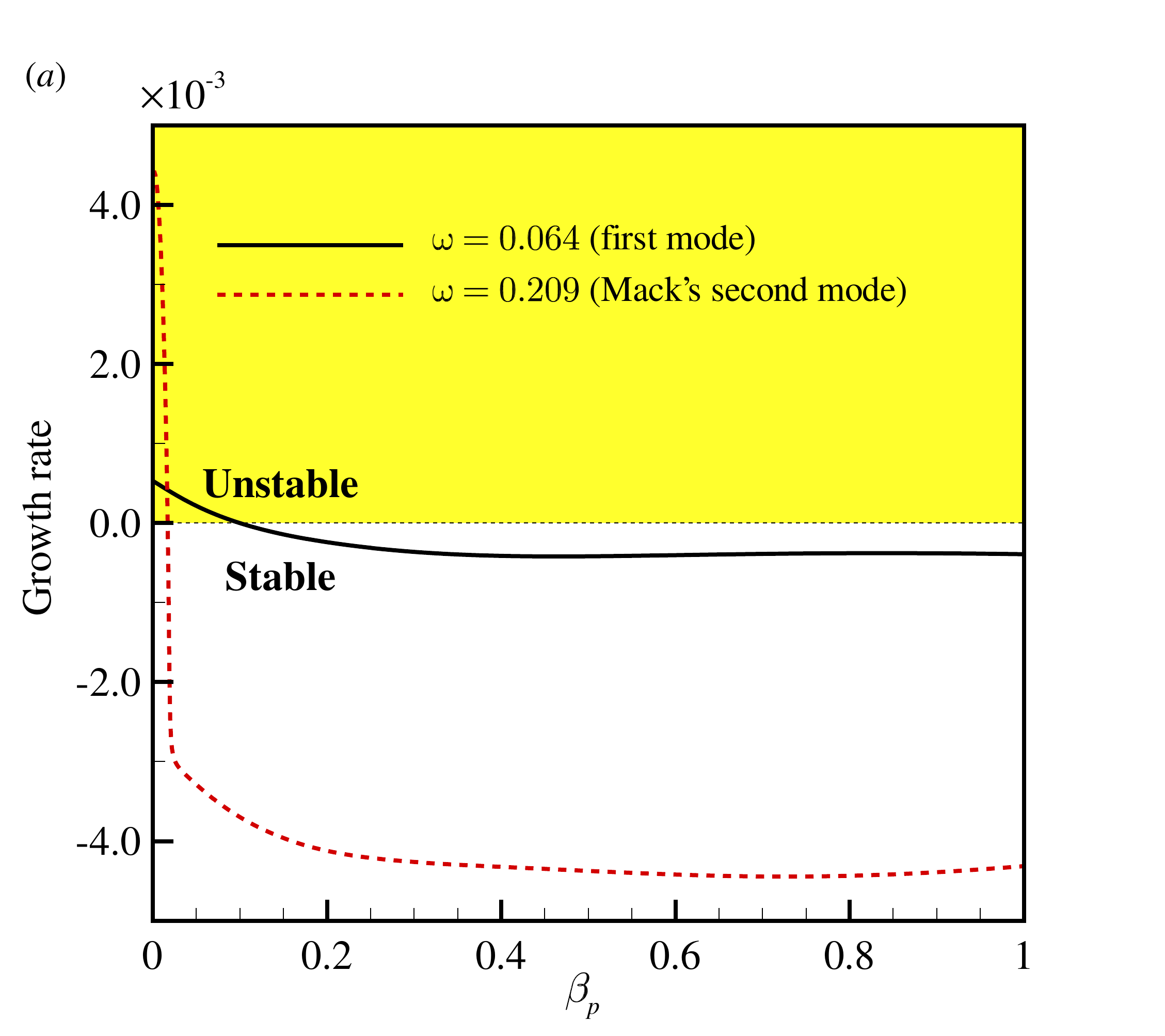}
    \includegraphics[width=0.48\linewidth]{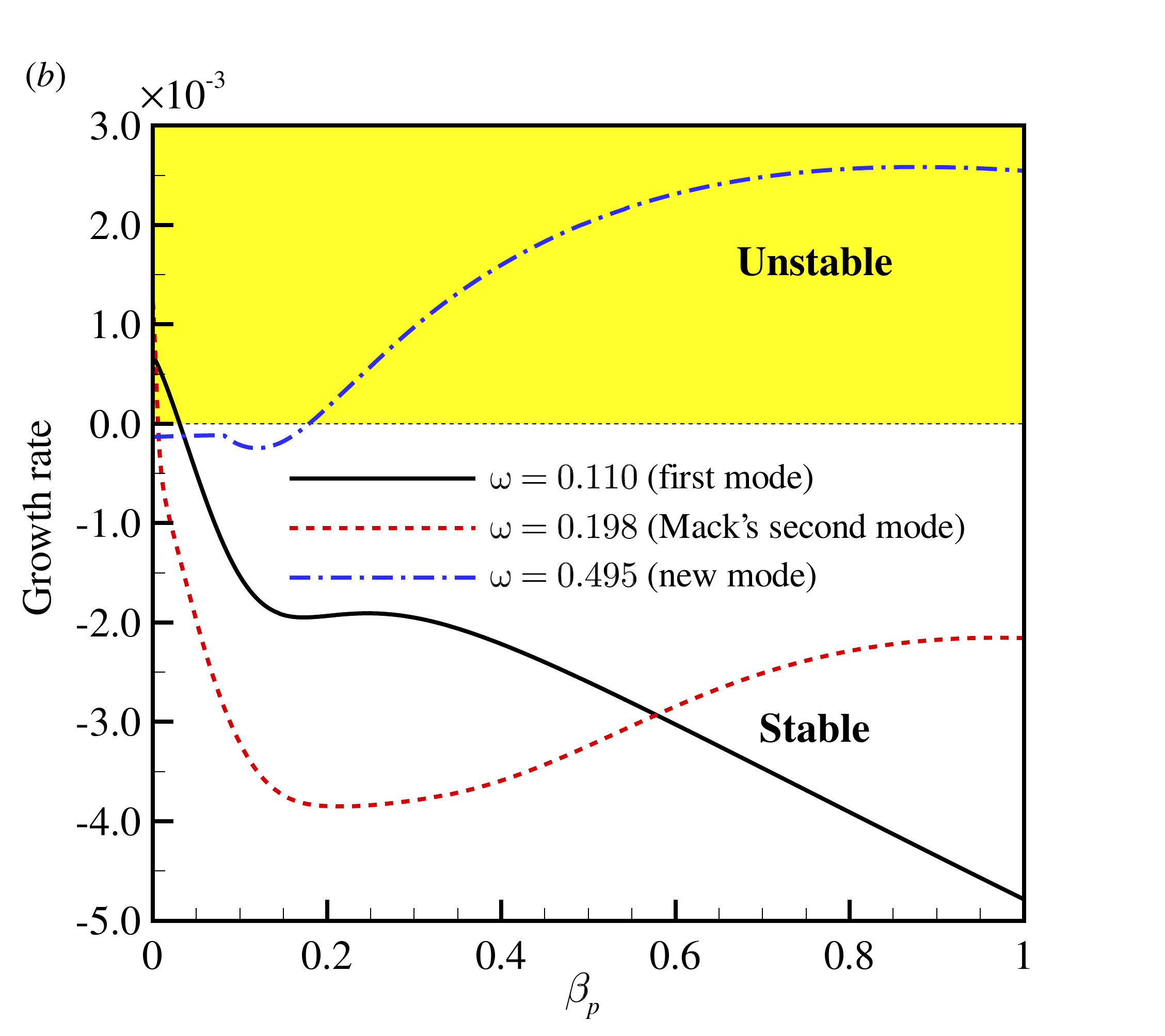}
    \caption{Growth rate of 2-D perturbations ($\beta=0$) as functions of pressure gradient $\beta_p$. (\textit{a}) Boundary layer with adiabatic condition ($H'_w$=0); (\textit{b}) Boundary layer with wall-heating ($H_w$=1.5);}
    \label{F3}
    \end{figure}

It is remarkable that the new mode becomes the only unstable modes when dual effects (wall-heating and FPG) are present. This phenomenon is interpreted in Figure~\ref{F3} where the influence of pressure gradient $\beta_p$ is revealed. Several representative frequencies were chosen to show the possible unstable modes. The results show that, whether the wall is heated or adiabatic, it does not change the significant stabilizing effect of FPG on the conventional modes. Both the first mode and Mack's second mode soon become stable when FPG increases to $\beta_p=0.1$. Mack's second mode is even more sensitive to this parameter. This is consistent with previous studies as introduced in Section~\ref{S1}. On the other hand, the new mode starts growing when $\beta_p$ reaches a value of about 0.2. Hence, the new mode becomes the only unstable mode in the boundary layer with FPG \& wall-heating.

\begin{figure}[H]
    \centering
    \includegraphics[width=0.48\linewidth]{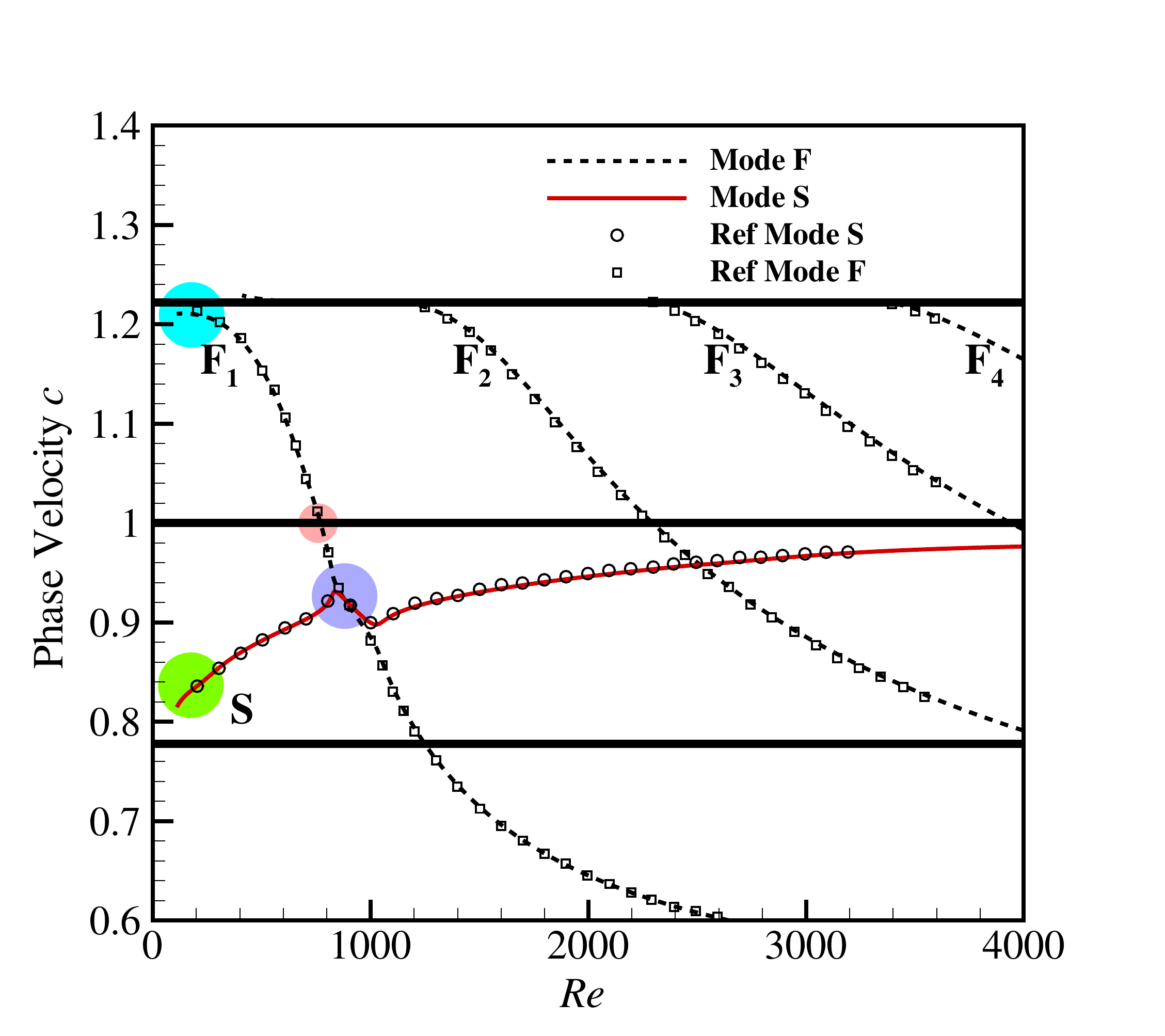}
    \includegraphics[width=0.48\linewidth]{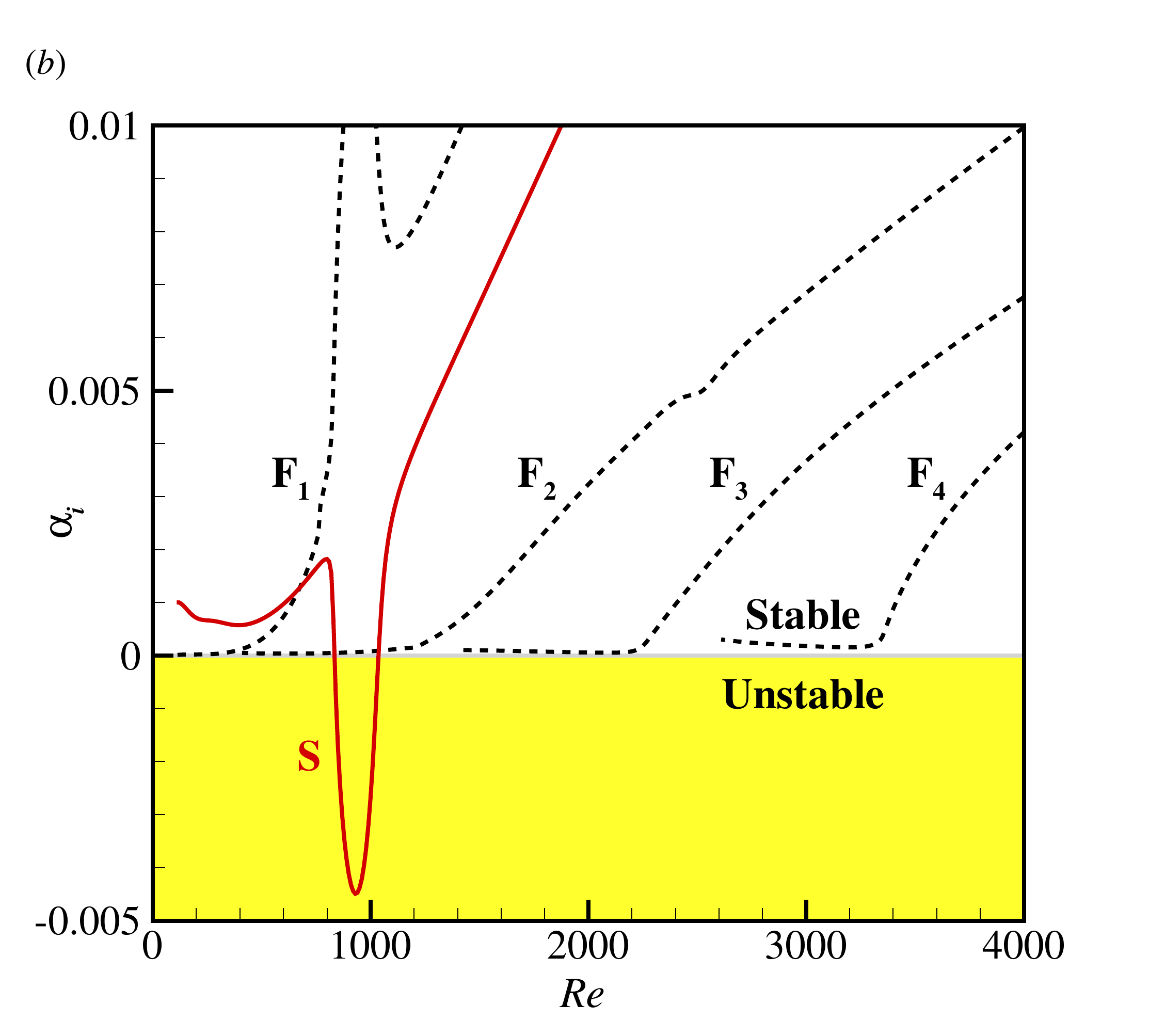}\\
    \includegraphics[width=0.48\linewidth]{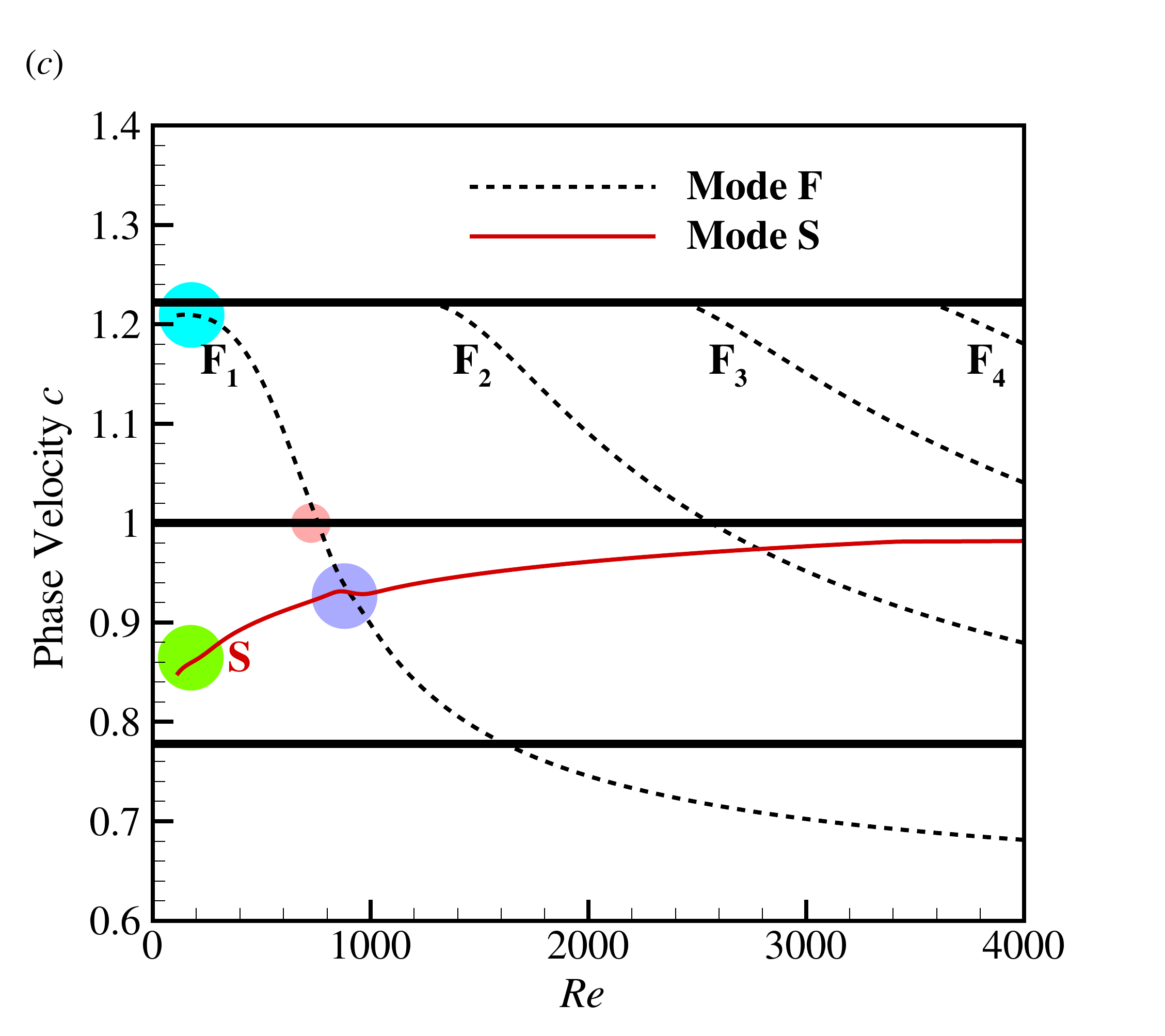}
    \includegraphics[width=0.48\linewidth]{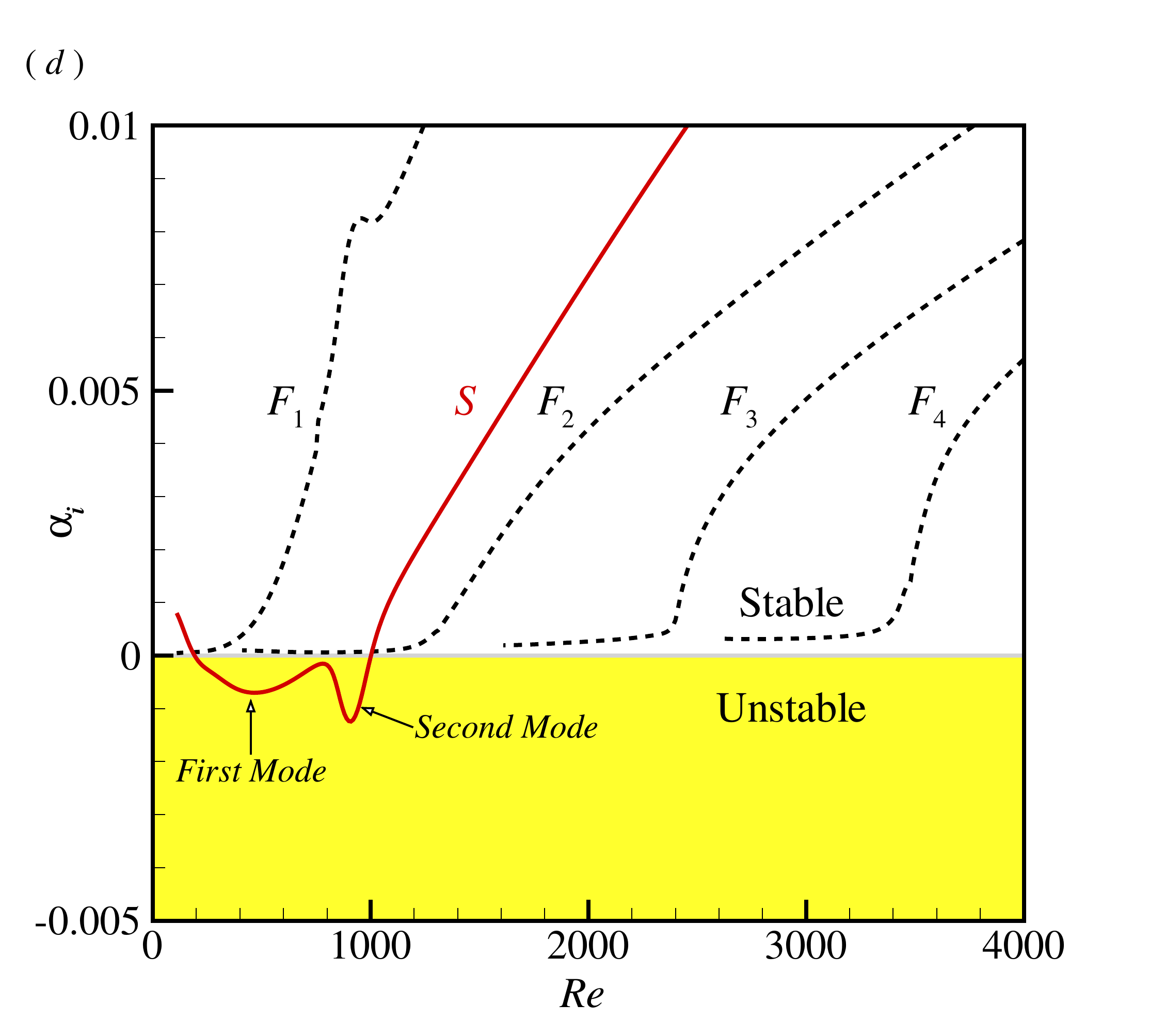}\\
    \includegraphics[width=0.48\linewidth]{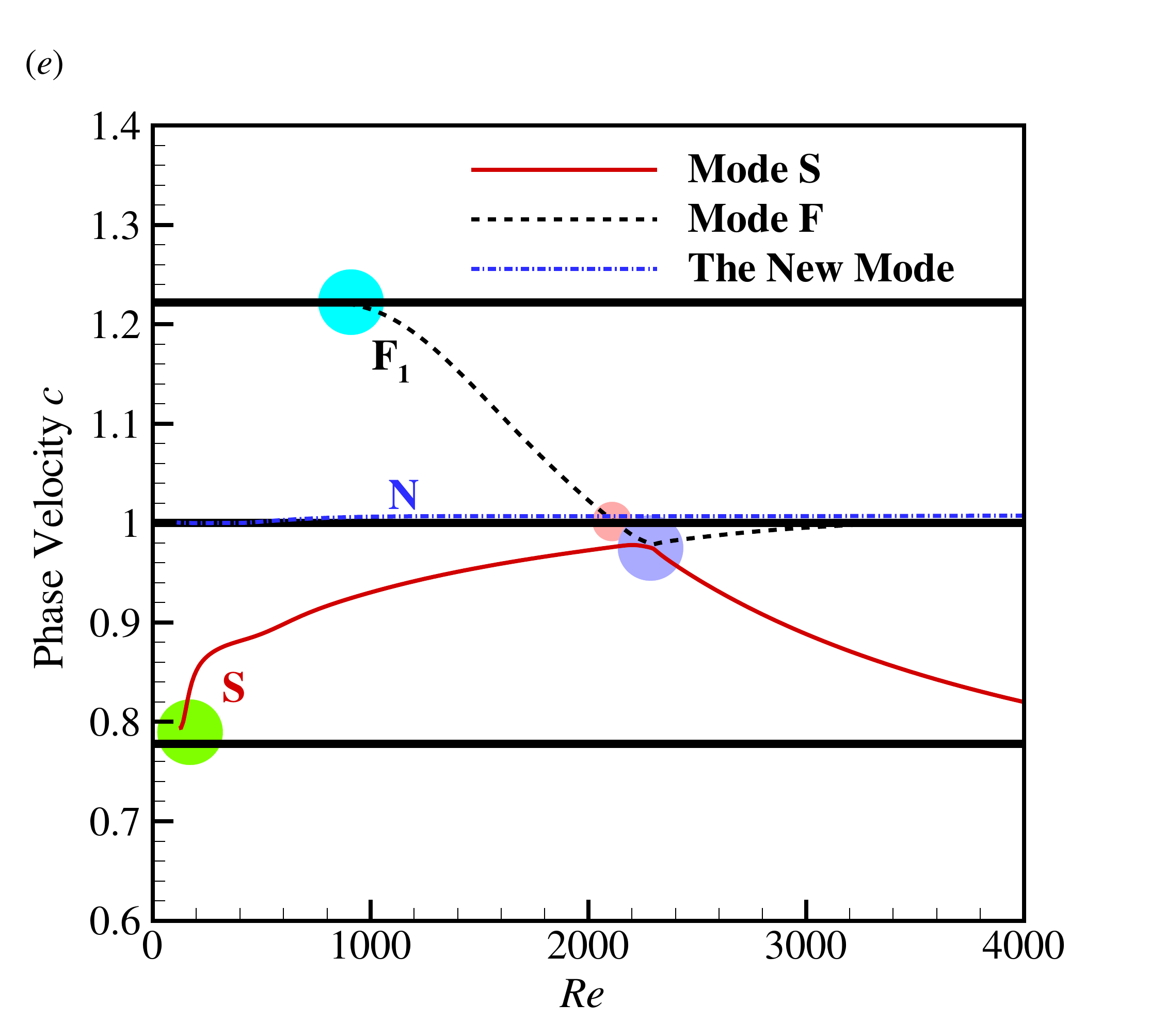}
    \includegraphics[width=0.48\linewidth]{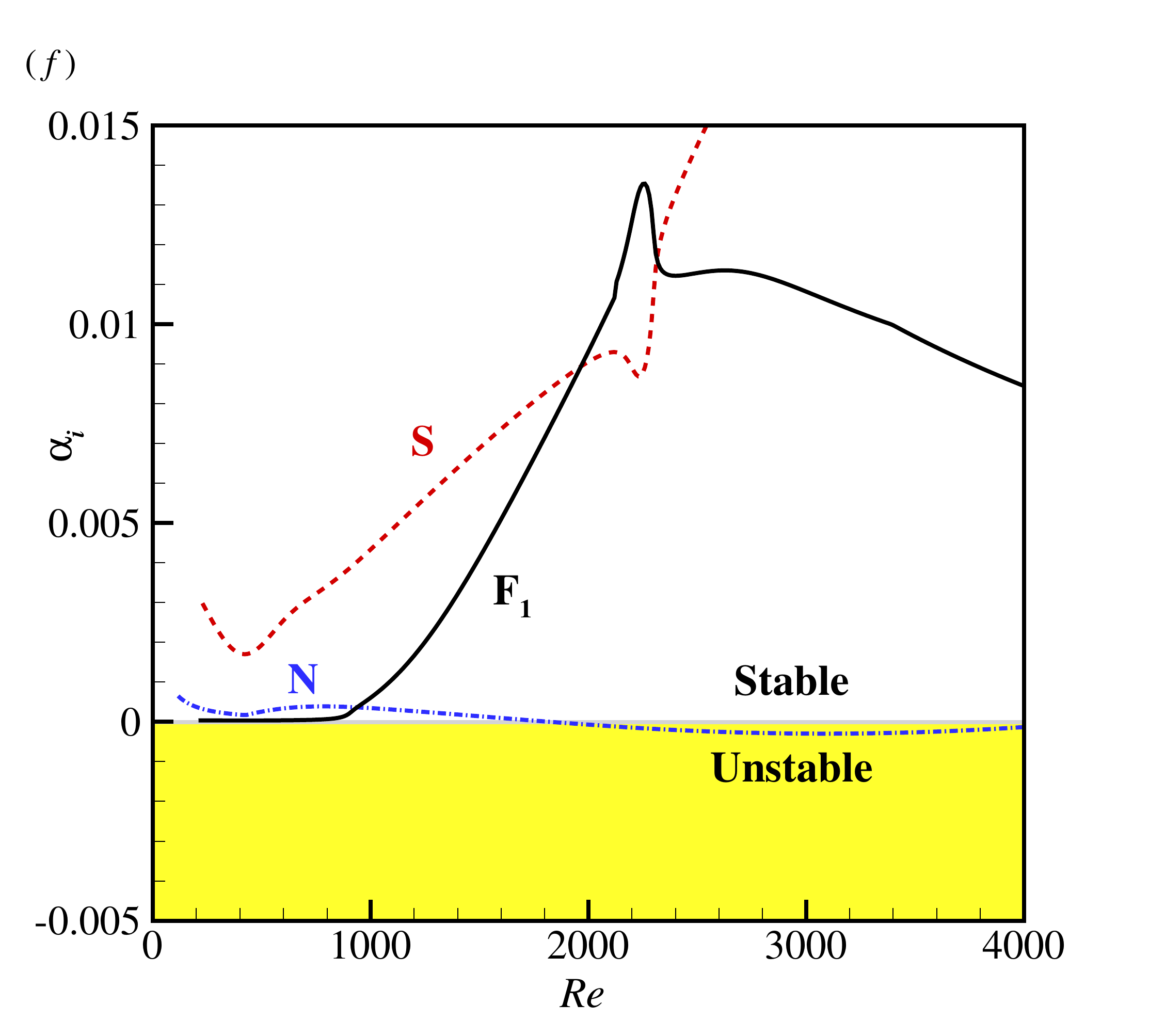}
    \caption{Spectrum of the 2-D perturbations ($\beta=0$) with frequency  $F=\omega/\Rey=2.2\times10^{-4}$ for Case 1 (\textit{a,b}); Case 2 (\textit{c,d}) and Case 3 (\textit{e,f}). The thick horizontal lines (in a,c,e) show the phase velocities of the continuous spectrum: fast acoustic wave ($c=1+1/\Ma$=1.22), voticity/entropy wave ($c=1.0$) and slow acoustic wave ($c=1-1/\Ma$=0.78).
    The circles show the synchronization regions. The notations $F_1$, $F_2$... represent the multiple Fast mode excited consecutively. The symbols in (\textit{a}) show the results from \cite{Ma2003A}}
    \label{F4}
    \end{figure}

The spectrum of high-speed boundary layers has been shown (see reviews by \cite{Fedorov2011,Zhong2012}) to help the understanding of the excitation of the unstable
modes. The synchronization between Mode F (stems from the fast acoustic wave) and Mode S (stems from the slow acoustic wave) gives rise to the growth of Mack's second mode. Detailed comments on the synchronization were made by \cite{Fedorov2011B}. Figure~\ref{F4} shows the discrete spectrum (phase velocity $c=\omega/\alpha_r$ and imaginary part of the eigenvalue $\alpha_i$.) for the three cases at fixed physical frequency $F=\omega/\Rey=2.2\times10^{-4}$.

Case 1 reproduced the spectrum in conventional adiabatic boundary layers with zero pressure gradient. At this frequency, Mode S played the first mode and Mack's second mode at different sections of Reynolds numbers. The first mode is stable while Mack's second mode enters the growth zone when Mode F and Mode S have almost identical phase velocities (synchronization). When the synchronization is finished (at about $\Rey=1200$), all the discrete modes decay. Case 2 is similar to Case 1 except the first mode has an unstable section due to wall-heating. The second mode, on the contrary, is stabilized by manifesting in a reduced overall growth rate.

The new mode appears in Case 3. One can still identify the Mode F and Mode S. However, Mode F synchronizes with the fast acoustic wave at a much larger Reynolds number. Synchronization between Mode F and Mode S still caused localized peak values of $\alpha_i$ for each other. Apparently, both modes are far from the unstable half-plane. Interestingly, the spectrum branching occurs indicating Mack's second mode has similar dispersion relation in this case. The new mode seems to stem from the vorticity/entropy wave ($c=1.0$) and remains a phase velocity slightly larger than 1.0. At $\Rey=1826$, the new mode becomes unstable and gains maximum growth rate at $\Rey=3090$.

    \begin{figure}[H]
    \centering
    \includegraphics[width=0.55\linewidth]{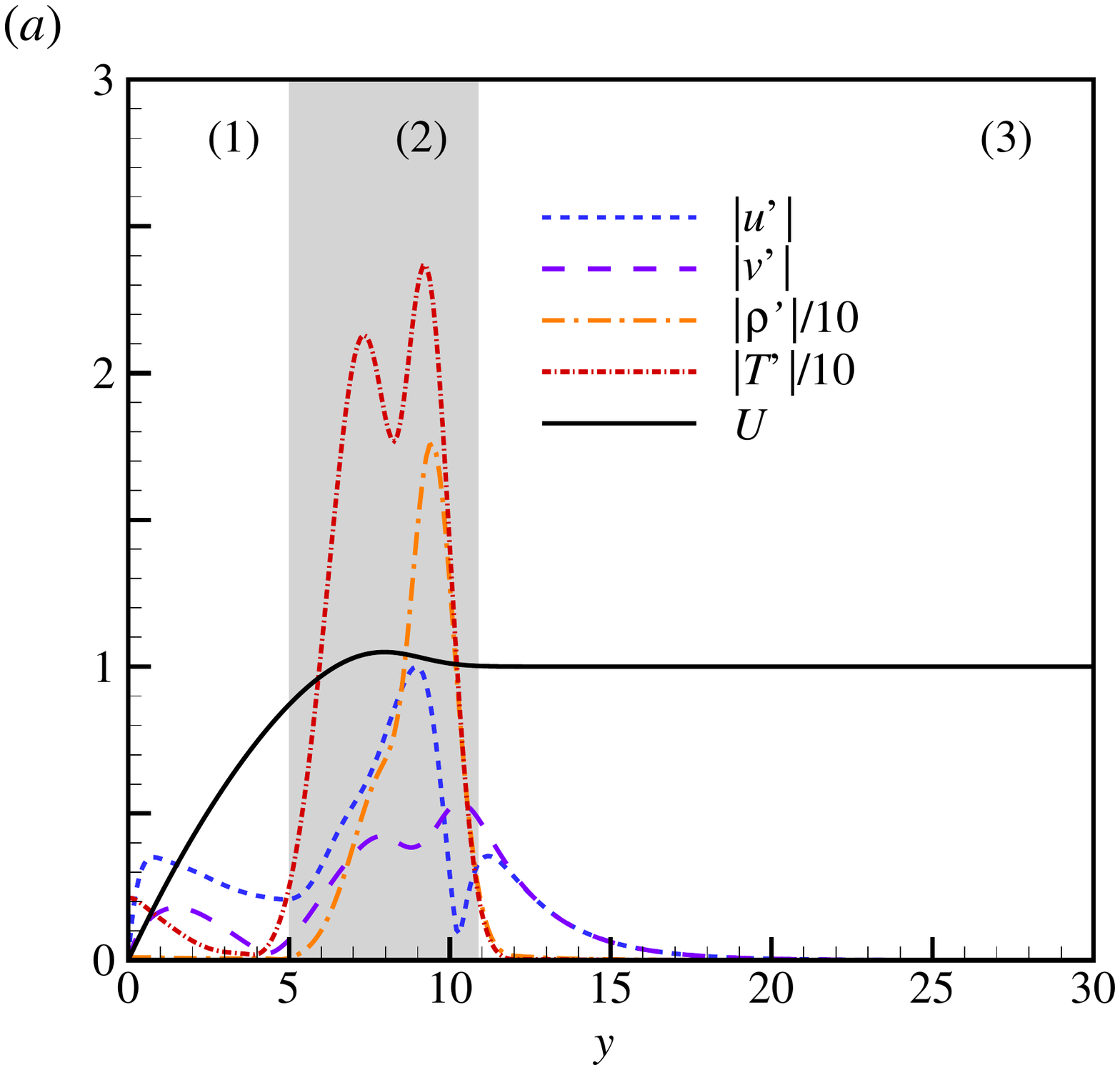}\\
    \includegraphics[width=0.49\linewidth]{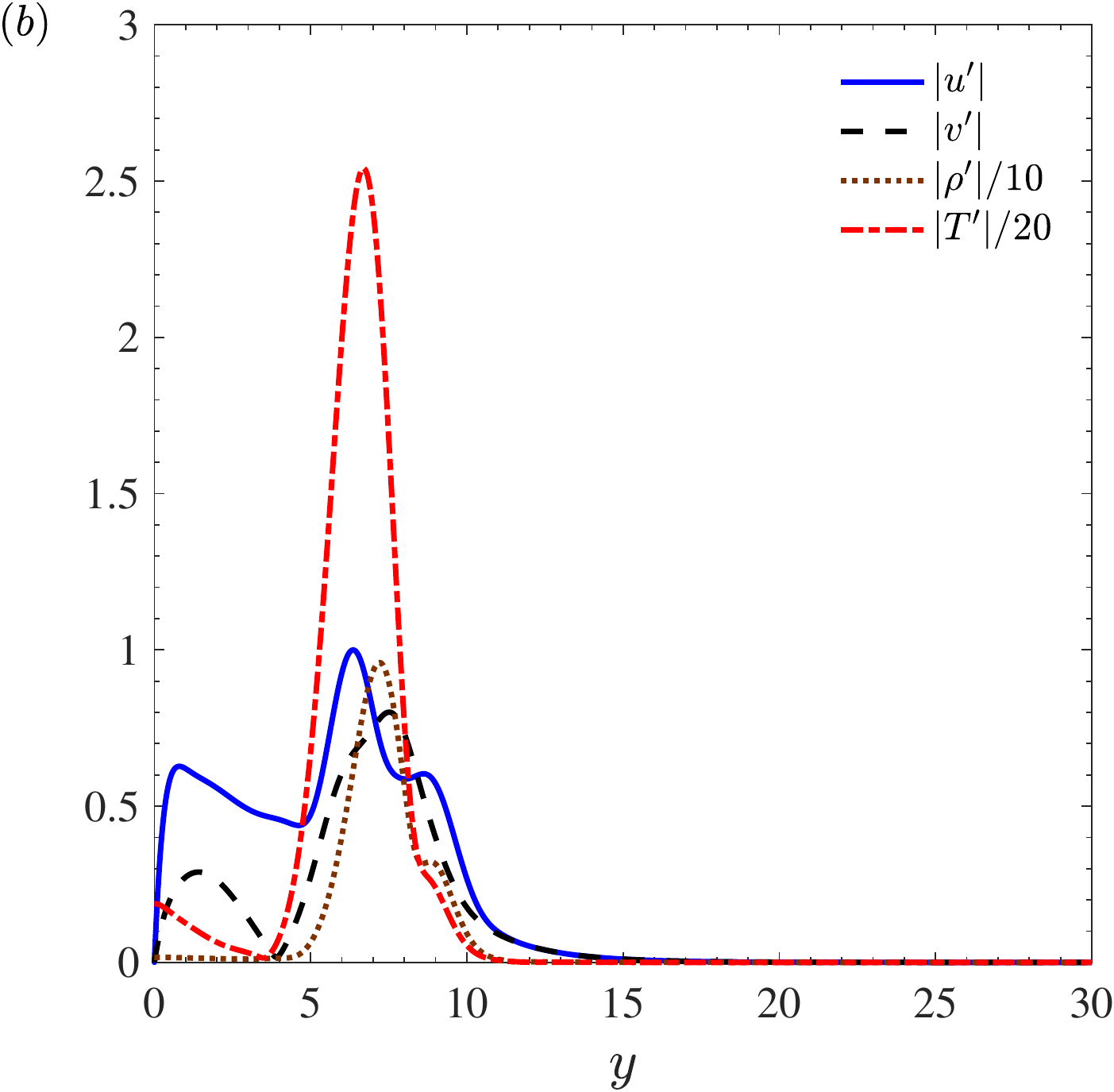}
    \includegraphics[width=0.49\linewidth]{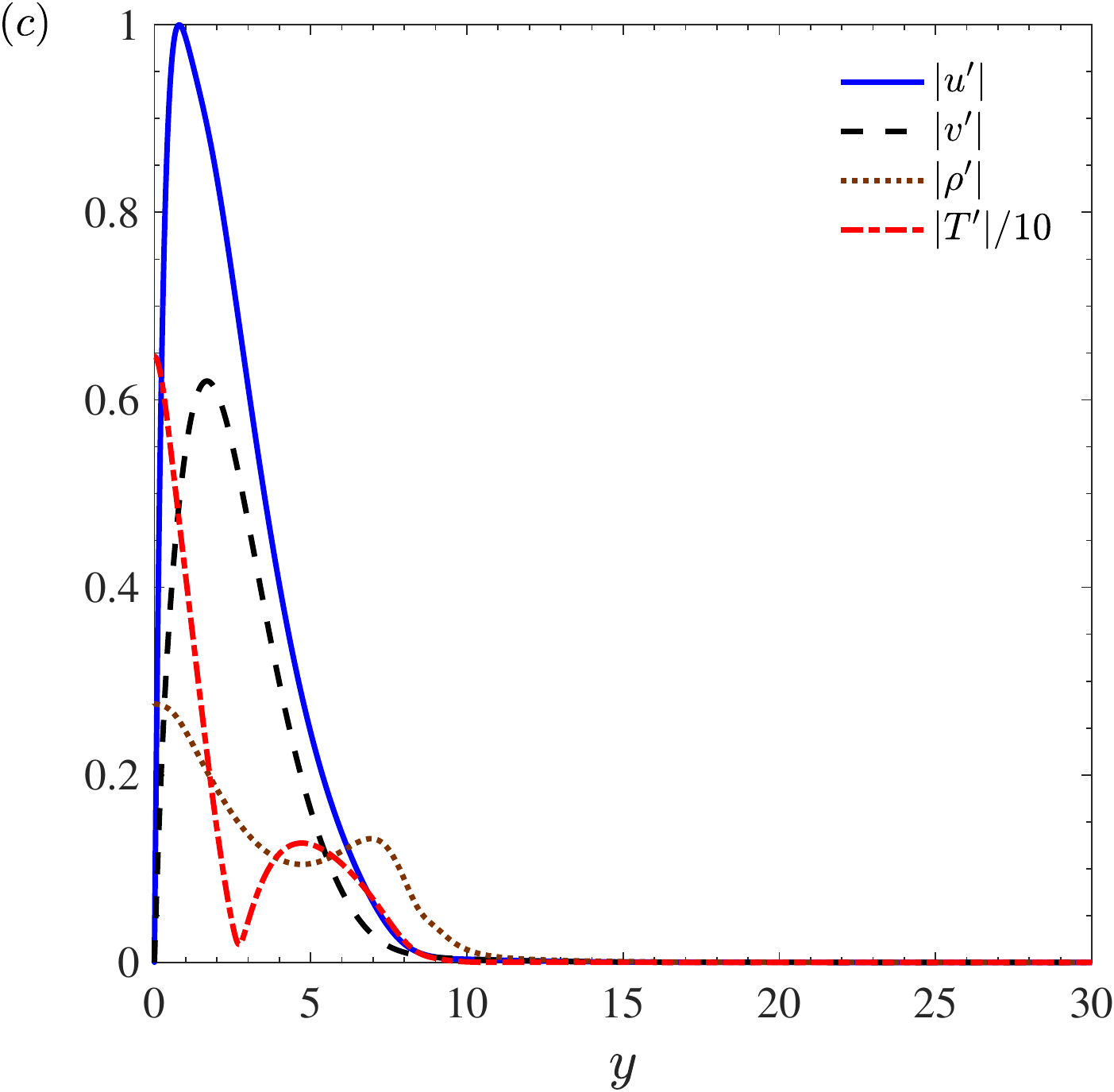}    
    \caption{Eigenvector of the new mode(figure (a)), mode S(figure (b)) and mode F(figure (c)) at $\Rey=2000$, $\omega=0.53$ and $\beta=0$. Absolute values are shown. The temperature and density perturbations are scaled with factors indicated in the legend.}
    \label{F5}
    \end{figure}

The eigenvectors of the most amplified new mode from Case 3 are shown in Figure~\ref{F5}a. Reynolds number $\Rey=2000$, $\omega=0.53$ and $\beta=0$. The base flow is plotted as a reference. The boundary layer can be qualitatively divided into three regions shown in the figure. Region (1) starts from the wall and is replaced by Region (2) where the overshoot $U(y)>1$ starts. Region (3) is the inviscid region outside the boundary layer. The perturbations are mainly distributed in Region (1) and (2) where the baseflow shear exists. As expected in most hypersonic cases, temperature perturbation has the maximum amplitude, followed by the density components. Both perturbations are largely distributed in Region (2). The velocity perturbation $\tilde{u}$ and $\tilde{v}$ though have much smaller amplitudes but are critical for the transportation of momentum and energy of the fluids. Besides, we show the eigenvector of the mode S and F in Figure~\ref{F5}b and Figure~\ref{F5}c for comparison. The perturbations of the mode F mainly locate near the wall while mode S becomes significant near the boundary layer edge.

\begin{figure}
    \centering
    \includegraphics[width=0.50\linewidth]{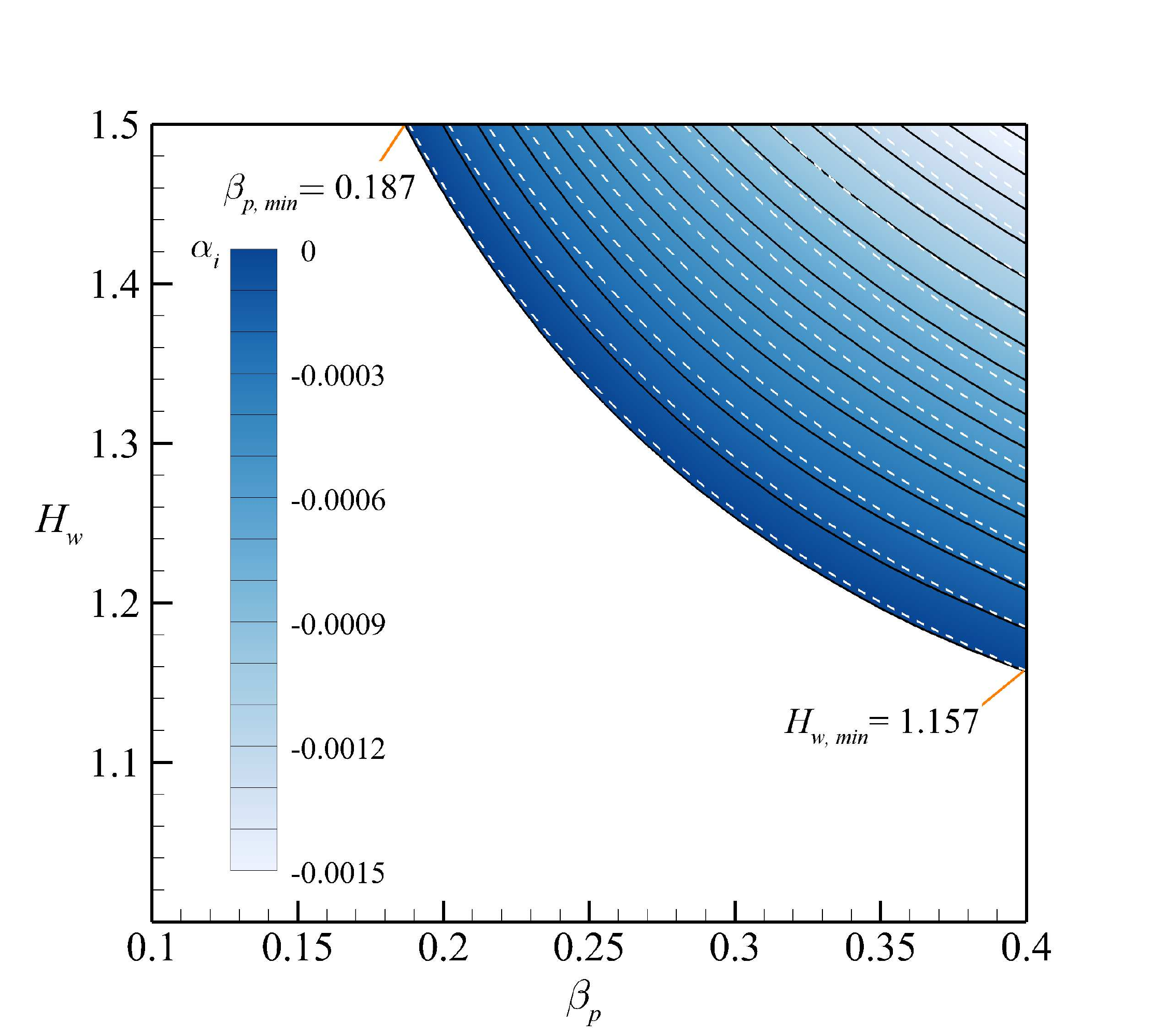}
    \caption{Isolines of the growthrate of the new mode at levels $-0.015\leq\alpha_i\leq0$ ($\alpha_i=0$ indicates the neutral curve). $\Rey=2000$, $\omega=0.53$, $\beta=0$, $1.0\leq H_w\leq1.5$, $0.1\leq\beta_p\leq0.4$. Solid (black) lines for $T_0^*=329K$ and dashed (white) lines for $T_0^*=1094K$.}
    \label{F6}
\end{figure}

To evaluate the potential importance of the new mode in practical flows. We pick the most unstable mode from Case 3 and study the influence of $H_w$ and $\beta_p$. Calculations are conducted with the total temperature of $T_0^*=329K$ and $1094K$ respectively. The corresponding free-stream temperatures $T_\infty^*=65.15K$ and $216.66K$ mimic the conditions of low-enthalpy experiment and flight at the altitude of 11km to 20km. As can be inferred from Figure~\ref{F6}, the effects of wall-heating and pressure gradient are complementary with regard to the growth of the new mode. Under flight conditions, the growth rate is slightly smaller. In both cases, the required minimal wall enthalpy $H_{w,min}=1.157$ and the pressure gradient $\beta_{p,min}=0.187$. This implies that the new mode has to meet severe conditions to become unstable. Particularly, the wall must be heated with additional sources and the pressure gradient should be large enough. Under flight conditions where the wall temperature can not exceed the adiabatic value, therefore, the new mode has no chance to appear. On the other hand, the new mode could be reproduced under experimental (artificial) conditions where new transition scenario shall be anticipated.

\section{Concluding remarks}\label{S4}
Inspired by Tunney \etal \cite{Tunney2015}, the viscous instability of the high-speed boundary layer with the dual effects of favorable pressure gradient (FPG) and wall-heating is studied. From modal stability analysis, the full stability diagram (in the coordinates of $\Rey-\beta-\omega$) is given and compared with conventional first mode and Mack's second mode. The new mode becomes the only unstable modes in such flows where FPG readily suppressed the conventional modes. The synchronization between the spectrum found in high-speed flows \cite{Fedorov2011B} remains but is not responsible for the growth of the new mode. Due to the requirement for additional heating, the new mode can be important only under experimental (artificial) conditions.

\section*{Acknowledgments}
The authors acknowledge the financial support by the National Natural Science Foundation
of China (Grant Nos. 11602127 and 11572176) and China Postdoctoral Science Foundation (No. 2016M590091 and No. 2017T100067).


\begin{thebibliography}{99}

\bibitem{Cebeci1974}
T.~Cebeci and A.~M.~O. Smith,
  { Analysis of turbulent boundary layers}.
  Academic Press, 1974.
  
\bibitem{Mack1984}
L.~M. Mack,
  Boundary-layer linear stability theory.
  { Special Course on Stability and Transition of Laminar Flow,
  AGARD-R-709}, 1984.

\bibitem{Mack1987}
L.~M. Mack,
  { Review of Linear Compressible Stability Theory}, pages 164--187.
  Springer New York, New York, NY, 1987.

\bibitem{Malik1989}
M.~R. Malik,
  Prediction and control of transition in supersonic and hypersonic
  boundary layers.
  { AIAA Journal}, 27(11):1487--1493, 1989.  

\bibitem{Kloker1990}
M.~Kloker and H.~Fasel,
  { Numerical Simulation of Two- and Three-Dimensional Instability
  Waves in Two-Dimensional Boundary Layers with Streamwise Pressure Gradient},
  pages 681--686.
  Springer Berlin Heidelberg, Berlin, Heidelberg, 1990.

\bibitem{Zurigat1992}
Y.~Zurigat, A.~Nayfeh, and J.~Masad,
  Effect of pressure gradient on the stability of compressible boundary
  layers.
  { AIAA Journal}, 30(9):2204--2211, 1992.
  
\bibitem{Masad1994}
J.~A. Masad and Y.~H. Zurigat,
  Effect of pressure gradient on first mode of instability in
  compressible boundary layers.
  { Physics of Fluids}, 6(12):3945--3953, 1994.
  
\bibitem{Saric1994}
W.~S. Saric,
  G\"{o}rtler vortices.
  { Annual Review of Fluid Mechanics}, 26:379--409, 1994.

\bibitem{Reed1996}
H.~L. Reed, W.~S. Saric, and D.~Arnal,
  Linear stability theory applied to boundary layers.
  { Annual Review of Fluid Mechanics}, 28:389--428, 1996.
  
\bibitem{Bech1998}
K.~H. Bech, D.~S. Henningson, and R.~A. W.~M. Henkes,
Linear and nonlinear development of localized disturbances in zero
  and adverse pressure gradient boundary-layers.
  { Physics of Fluids}, 10(6):1405--1418, 1998.

\bibitem{Cebeci2002} 
T.~Cebeci,
  { Convective heat transfer}.
  Springer, 2002.

\bibitem{Ma2003A}
Y.~Ma and X.~Zhong,
  Receptivity of a supersonic boundary layer over a flat plate. part 1.
  wave structures and interactions.
  { Journal of Fluid Mechanics}, 488:31--78, 7 2003.
  
\bibitem{Saric2003}
W.~S. Saric, H.~L. Reed, and E.~B. White,
  Stability and transition of three-dimensional boundary layers.
  { Annual Review of Fluid Mechanics}, 35:413--440, 2003.

\bibitem{Ricco2009}
P.~Ricco, D.-L. Tran, and G.~Ye,
  Wall heat transfer effects on \textsc{K}lebanoff modes and
  \textsc{T}ollmien--\textsc{S}chlichting waves in a compressible boundary
  layer.
  { Physics of Fluids}, 21(2):024106, 2009.
  
\bibitem{Fedorov2011B}
A.~Federov and A.~Tumin,
  High-speed boundary-layer instability: Old terminology and a new
  framework.
  { AIAA Journal}, 49(8):1647--1657, 2011.

\bibitem{Fedorov2011}
A.~Fedorov,
  Transition and stability of high-speed boundary layers.
  { Annual Review of Fluid Mechanics}, 43:79--95, 2011.

\bibitem{Zhong2012}
X.~Zhong and X.~Wang,
  Direct numerical simulation on the receptivity, instability, and
  transition of hypersonic boundary layers.
  { Annual Review of Fluid Mechanics}, 44:527--561, 2012.
  
\bibitem{Franko2014}
K.~J. Franko and S.~Lele,
  Effect of adverse pressure gradient on high speed boundary layer
  transition.
  { Physics of Fluids}, 26(2):176--183, 2014.

\bibitem{RJ2014b}
J.~Ren and S.~Fu,
  Competition of the multiple \textsc{G}\"{o}rtler modes in hypersonic
  boundary layer flows.
  { SCIENCE CHINA Physics, Mechanics \& Astronomy},
  57(6):1178--1193, 2014.

\bibitem{RJ2015b}
J.~Ren and S.~Fu,
  Secondary instabilities of \textsc{G}\"{o}rtler vortices in
  high-speed boundary layer flows.
  { Journal of Fluid Mechanics}, 781:388--421, 2015.

\bibitem{RJ2016}
J.~Ren S.~Fu and A.~Hanifi,
  Stabilization of the hypersonic boundary layer by finite-amplitude streaks.
  {Physics of Fluids}, 28(2):024110, 2016.
  
\bibitem{Fedorov2015}
A.~Fedorov, V.~Soudakov, I.~Egorov, A.~Sidorenko, Y.~Gromyko, D.~Bountin,
  P.~Polivanov, and A.~Maslov,
  High-speed boundary-layer stability on a cone with localized wall
  heating or cooling.
  { AIAA Journal}, 53(9):2512--2524, 2015.

\bibitem{Tokugawa2015}
N.~Tokugawa, M.~Choudhari, H.~Ishikawa, Y.~Ueda, K.~Fujii, T.~Atobe, F.~Li,
  C.~L. Chang, and J.~White,
  Pressure gradient effects on supersonic transition over axisymmetric
  bodies at incidence.
  { AIAA Journal}, 53(12):3737--3751, 2015.

\bibitem{Tunney2015}
A.~P. Tunney, J.~P. Denier, T.~W. Mattner, and J.~E. Cater,
  A new inviscid mode of instability in compressible boundary-layer
  flows.
  { Journal of Fluid Mechanics}, 785:301--323, 2015.  
  
\bibitem{Costantini2016}
M.~Costantini, S.~Hein, U.~Henne, C.~Klein, S.~Koch, L.~Schojda, V.~Ondrus, and
  W.~Schr\"{o}der,
  Pressure gradient and nonadiabatic surface effects on boundary layer
  transition.
  { AIAA Journal}, 54(11):3465--3480, 2016.

\bibitem{BL}
H.~Schlichting and K.~Gersten,
  { Boundary-Layer Theory}.
  Springer Berlin Heidelberg, Berlin, Heidelberg, 2017.

\end{thebibliography}
\end{document}